# (Fe,Sn)$_4$N alloy as model spin-glass system with short-range competing interactions on a non-frustrated simple cubic lattice.

Sergii Khmelevskyi, and Peter Mohn.

Center for Computational Materials Science, Institute for Applied Physics, Vienna University of Technology, Wiedner Hauptstrasse 8, A-1040, Vienna, Austria.

The origin of the spin-glass state in (Fe,Sn)$_4$N alloys is studied on the basis of a Heisenberg Hamiltonian with parameters derived from first-principles within the magnetic force theorem applied in the framework of the Disordered Local Moments method and Local Spin Density Approximation. We show that in the alloy concentration range where the spin-glass state is stable only one Fe sub-lattice is intrinsically magnetic and the inter-atomic exchange magnetic interactions are essentially short-ranged due to effects of chemical and magnetic disorder. The magnetic Fe atoms with well localized spin-moments are randomly distributed over the non-geometrically frustrated simple cubic lattice. The magnetic frustration, which generally is believed to be an essential ingredient of the spin-glass state formation condition, may occur only due to the competition of the two nearest neighbor interactions. We thus argue that (Fe,Sn)$_4$N is a rare example of a spin-glass system where the mechanism of spin-glass state formation might be studied in the framework of the minimal random-site model on a simple cubic lattice with competing interactions, while the effects of the geometrical frustration can be excluded.

The spin-glass problem remains one of the most complicated topics in modern solid state physics.[1] Over more than four decades since the discovery of the phenomenon[2] in dilute magnetic alloys a vast amount of experimental data has been collected. The "experimental" definition of the spin-glass state and the spin-freezing process has been worked out in great detail.[3] However, the theoretical description of the spin glass (SG) state formation in real materials (*realistic spin glasses*)[4] is still elusive and remains one of the central unsolved problems in condensed matter physics.[5] In the past most of the effort has been concentrated on models describing "mathematical" spin glasses, which provide, as it is believed, a crucial simplification by replacing the chemical site-disorder in real alloys by bond-disorder on an ideal lattice.[6] Magnetic models with bond-disorder[7] allow for the celebrated *replica* symmetry trick, which has been successfully used to describe the spin-glass transition in both in Ising- and Heisenberg-type of spin systems.[8,9] The rigorous treatment of the problem, however, requires a consideration of the solutions with broken replica symmetry[10] and the associated complexity has given rise to the brilliant mathematical theory[11] of the random-bond spin-glass, which is still under intensive development.[12] Surprisingly, much less theoretical attention has been paid to the realistic SG alloys, which are random-site disordered systems on a lattice. Even large scale computational computer simulations have concentrated mainly on the mathematical glasses, in attempt to solve the associated theoretical problems of random-bond models, rather than on the random-site SGs.[13] This has led to a rather interesting dichotomy in the experimental and theoretical spin glass research, mentioned already in 90ies[4] when experimental works dealing with realistic glasses refer to the results of mathematical spin-glass theory and *vice-versa* theoretical works on mathematical glasses refer to experimental results for the random-site SG alloys for illustrations and comparison. The root of this dichotomy is the above mentioned development of the mathematical SG theory, which provides a wide speculative background for discussion, and presumably removes the ultimate complexity[4] of the SG problem on a lattice with site-disorder. What certainly can be learned from SG theory and empirical data up to date is that an essential ingredient of the SG state formation in real alloy systems is magnetic frustration and chemical disorder[3]. The frustration might have two sources – geometrical frustration of the lattice and/or the competition of the inter-site exchange interactions. In the most studied "canonical" 3 spin-glass systems, like **Cu**Mn, **Au**Fe and some other fcc alloys, both types of the frustration are present due to natural frustration of the fcc lattice with respect of antiferromagnetic interactions and the long range oscillating RKKY interaction.[1,3] In fact in realistic SG alloys other physical factors apart from frustration and disorder are debated to be important for the onset of a SG state such as the magnetic anisotropy,[14,15] the Dzyaloshinskii-Moriya exchange,[16] long-ranginess of the interactions (RKKY glasses)[3] and complex spin dynamics effects.[17] However, the problem whether these effects are necessary for the onset of a SG state in addition to frustration and disorder remains unclear. To resolve these issues it is desirable to consider realistic site-disorder magnetic models and to investigate the SG system behavior with realistic material specific parameters.

In the last decades highly efficient first-principles based computational quantum mechanical methods have been developed and applied to disordered magnetic alloy systems.[18,19] In particular, the methodology of the ab-initio calculations of the magnetic exchange interactions in the framework of the magnetic force theorem (MFT) embedded[20] in the Local Spin Density Approximation[21] (LSDA) are very successful in the prediction of the complex ground states and magnetic ordering temperatures of the metallic magnets.[22] It has also been demonstrated that first-principles based MFT is able to account for atomic disorder and chemical substitution effects on the exchange interactions in a number of alloy systems allowing for quantitative analysis of such effects as the evolution of the magnetic ground state with non-magnetic substitutions,[23] the change of the magnetic critical temperature due to partial disorder effects,[24] and the prediction of the character of the meta-magnetic processes in alloys with non-collinear magnetic structure.[25] The first-principles estimation of the exchange interactions in real SG alloys can be traced back to the pioneering work by Ling *et al.*[26] who calculated the interactions up to 9$^{th}$ nearest neighbor (NN) shells in $Mn_{15}Cu_{85}$. These long distance exchange interactions have been calculated on the basis of the LSDA-MFT for the **Au**Fe reentrant spin glass alloys in a wide range of chemical compositions.[27] It was shown,[27] in particular, that the antiferromagnetic coupling, which "frustrates" the magnetic system, appears only between very distant Fe atomic neighbors and that the simple RKKY form of the interaction, often used in model SG simulations,[28] is only approximately valid because of its strong dependence on the lattice directions, and the existence of an additional exponential decaying factor due to chemical and magnetic disorder, and a significant contribution from a direct exchange mechanism in the first few NN shells. Realistic exchange constants calculated from first-principles have been used in large scale Monte-Carlo simulations[29,30] and atomistic spin dynamic simulations[31] of the SG state in **Cu**Mn alloys. These studies allow to progress in the understanding of the SG formation and spin dynamics in realistic SG alloy systems. However, they also revealed additional complexities when dealing with canonical SG alloys. In particular, the presence of significant short-range chemical order, which has an impact on the magnetic properties of the **Cu**Mn alloys.[26,30] The most studied canonical[3] SG alloys, like **Au**Fe and **Cu**Mn, have geometrically frustrated face-centered cubic (fcc) structure. It means that both types of magnetic frustration are present in fcc alloys: the geometrical one and the frustration due to a competition of the distant exchange interactions. Thus, the theoretical simulations of canonical fcc SG alloys, even with realistic parameters, always leaves open the question concerning the minimal set of conditions for a SG state formation. In this regard a recent experimental report[32] on the spin-glass behavior in $(Sn_{0.9}Fe_{3.1})_4N$ alloy is rather transparent since, as we will advocate in this work on the basis of the first-principle simulations, it appears that the SG state is forming in this material on the non-geometrically frustrated simple cubic (sc) lattice. By calculating the magnetic moments and exchange interactions in the high-temperature paramagnetic state in chemically disordered alloys we show that only Fe moments on one of the two iron sub-lattices are intrinsic in $(Sn_{0.9}Fe_{3.1})_4N$, where this sub-lattice is essentially atomically disordered and the inter-site exchange interactions are short ranged. These observations makes $(Sn_{0.9}Fe_{3.1})_4N$ to be an almost ideal model system for a combined experimental and model study of SG phenomena on a non geometrically frustrated simple cubic lattice with random site disorder. The frustration arises in this system entirely due to competition of the first and second nearest neighbor interactions, which are both antiferromagnetic and have a similar magnitude. Quite notoriously that the Heisenberg model on the simple cubic lattice without site-disorder and competing 1$^{st}$ and 2$^{nd}$ nearest neighbor interactions ($J_1$-$J_2$) has been considered as a simplest "canonical" 3D-model for a study on the emergence of a spin liquid behavior and quantum fluctuations due to frustration of the interactions.[33] The possible relevance of the $J_1$-$J_2$ antiferromagnetic Heisenberg model on a sc lattice with site disorder to the SG problem, however, remains an open question.[34] The experimental observation[32] of the SG state in $(Sn_{0.9}Fe_{3.1})_4N$ together with a structure of its paramagnetic state and exchange interactions revealed in the present work may be stimulating for further discussions and modeling of the spin glass problem in the framework of this simple (perhaps the simplest realistic) model of a disordered magnetic alloy.

The parent compound $Fe_4N$ crystallizes in the cubic anti-perovskite crystal structure ($Pm\bar{3}m$, No. 221, see figure 1) and its magnetic properties have been rather well studied experimentally[35,36,37] as well as theoretically on a first-principles basis.[38,39,40] The strong ferromagnetic interactions between two Fe sub-lattices, Fe(1a) and Fe(3c), in $Fe_4N$ resulting in a remarkable total magnetization and high magnetic ordering temperature (767 K).[41] The exchange interactions between the Fe spin moments in the ferromagnetic ground state has been studied in the ab-initio framework by Meinert *et al.*,[40] who found a rather reasonable agreement between the calculated and experimental Curie temperature. In this work it has also been found that besides of a strong ferromagnetic nearest neighbor interaction there are antiferromagnetic distant interactions of considerable magnitude. Based on Meinert's results[40] and their own ab-initio calculations, which show the overall weakening of the ferromagnetism upon Sn substitution in

Fe$_4$N, Scholz and Dronskowski[32] interpreted the formation of the SG state in (Sn$_{0.9}$Fe$_{3.1}$)$_4$N alloy as a result of the competition between the weakened ferromagnetic (FM) and the distant antiferromagnetic (AFM) interactions. In this work, however, we show that indeed only Fe in position 1a (see figure 1) possesses intrinsic magnetic moments in (Sn$_{0.9}$Fe$_{3.1}$)$_4$N and that these moments are randomly distributed over the sc lattice (with 10% sites occupied) interact antiferromagnetically. We thus argue that the SG state is formed due to competitions of the two NN AFM couplings on the site-disordered sc lattice and support our argumentation by Monte-Carlo simulations with ab-initio calculated exchange interactions.

To this end we performed electronic structure calculations for disordered (Fe$_{4-x}$Sn$_x$)N alloys in the concentration range x=0-0.9 employing the Coherent Potential Approximation (CPA) in the framework of the Korringa-Kohn-Rostoker (KKR) method and the Atomic Sphere Approximation (ASA).[42,43] In our KKR-ASA calculations the partial wave functions were expanded in a spdf-basis (up to $l=3$) and the effects of exchange and correlation are treated within the Local Spin Density Approximation (LSDA).[44]

The magnetic exchange interaction constants, $J_{ij}$, of the classical Heisenberg Hamiltonian:

$$H = -\sum_{i,j} J_{ij} \vec{e}_i \vec{e}_j \quad (1)$$

where $\vec{e}_i$ is the unit directional vector of the magnetic moment at the $i$-th site of the Fe sublattice, have been calculated using the magnetic force theorem[45] as implemented in the bulk Korringa-Kohn-Rostoker band structure method.[46] In fact, our calculation methods are similar to those used by Meinert[40] for Fe$_4$N and our exchange constants calculated for ferromagnetic state of stoichiometric Fe$_4$N (see upper panel of the figure 2) are very similar as well. However, our discussion in this work will be based on the exchange constants calculated in the paramagnetic state with disordered local moments (DLM)[47] above the magnetic ordering temperature. The importance of the paramagnetic state as a reference state for calculations of the exchange interactions in metals for investigations of the high temperature properties and, in particular, of the magnetic phase transition has been pointed out and discussed several times.[48,49] One must also take into account that for $(Sn_{0.9}Fe_{3.1})_4N$, which experimentally does not show long range order (LRO) the choice some particular ordered magnetic configuration for the estimation of the exchange constants would be rather artificial and in general might lead to wrong conclusions. The thermal magnetic disorder essentially modifies the electronic structure of the valence bands of metals and consequently the exchange interactions, an example will be given just below for stoichiometric $Fe_4N$. The use of the classical Heisenberg model for metallic systems is dictated by the band origin of the atomic magnetic moments. Further discussion on the application of the classical Heisenberg model and its extensions for investigation of finite temperature magnetic properties of 3d-metallic systems can be found, e.g., in our recent work.[50]

In figure 2 we show the exchange constants of the Hamiltonian (1) calculated in the FM ground state and the DLM state above $T_c$ for $Fe_4N$. We see that the dominating interaction is the 1$^{st}$ NN inter-sublattice ferromagnetic coupling between Fe atoms in position 1a and 3c. This interaction defines the ferromagnetic order at low temperatures. One immediately notes the difference between the interactions calculated in the FM and DLM state. In the FM state they are essentially long range with a few AFM intra-sublattice (1a-1a as well as 3c-3c) couplings of rather significant sizes. In the DLM state the distant interactions are damped by the magnetic disorder effects on the electronic structure. If one wants to discuss low temperature properties (much lower than the magnetic ordering temperature), like spin-waves or magnetization dynamics etc, one needs to consider the exchange interactions calculated in the FM state. But if we want to study the onset of magnetic order at high temperatures, than the DLM interactions are relevant. Only for magnetic insulators with fully localized magnetic moments the interactions in the FM and paramagnetic state with DLM might be the same. In metals, where the interactions are mediated by the metallic bonds, the magnetic disorder effects will always modify the inter-atomic exchange interactions via changes of the electronic structure of the conduction bands. It has been shown that even in the case of well localized systems like hcp Gd these effects are important.[48] In the paramagnetic state only three interactions are essential for the onset of the magnetic order in $Fe_4N$ – strong 1NN coupling between 1a and 3c sub-lattices, and two (1NN and 2NN) interactions with opposite sign within the 2a sub-lattice. Estimating the magnetic ordering temperature by performing the Monte-Carlo simulations for the Hamiltonian (1) using the calculated exchange energies (lower panel of figure 2) we obtain a value of 790 K, which is in good agreement with experiment (767 K).[41] In these simulations we used interactions up to the 15NN shells for each of the Fe sub-lattices. These results justify our further discussion of the exchanges in the $(Sn_{0.9}Fe_{3.1})_4N$ alloys using DLM approach to the electronic structure of the paramagnetic state. Before we proceed we note here that our exchanges calculated for the FM state are similar to those obtained previously by Meinert[40] for stoichiometric $Fe_4N$.

The considerable difference in the magnitude of the interactions for the FM and DLM states of $Fe_4N$ can be further understood by analyzing the atomic magnetic moments on both Fe sub-lattices (see figure 3a). The moment of Fe in position 1a is rather large (~3 $\mu_b$), having the same value in the FM ground state and in the paramagnetic state suggesting a high degree of localization. The moment of Fe in the position 3c is smaller - in FM ground state it is almost exactly 2 $\mu_b$ - but it largely reduces in the paramagnetic DLM state to 1.24 $\mu_b$. This is a signal for the very itinerant character of the magnetism on the Fe (3c). Experimentally it is known that upon doping the Sn atoms substitute the Fe atoms entirely in the 1a position.[32] Our CPA calculations suggest a gradual collapse of the Fe moments in 3c position (figure 3a) in both FM and DLM states with increasing Sn concentration. At the critical Sn concentration in $(Fe_{4-x}Sn_x)N$ alloy (approximately at x=0.2) Fe(3c) loses its intrinsic moment in the paramagnetic state which, however, might be induced upon ferromagnetic ordering of Fe in 1a position at low temperatures. Due to this effects the Fe(1a)-Fe(3c) FM interaction weakens very fast with increasing Sn concentrations (see the respective calculated values in figure 3b) and already at $x \approx 0.2$ it has almost no influence on the magnetic order formation at elevated temperatures. The leading interactions, which determines the magnetic ordering in $(Fe_{4-x}Sn_x)N$ alloys with high Sn concentration would be the interactions between large localized Fe moments in 1a positions. Quite an interesting observation is that the magnitude of these interactions (1a-1a) dramatically changes with increase of the Sn concentration. The 1NN 1a-1a interactions being ferromagnetic in stoichiometric $Fe_4N$ compound turn to be antiferromagnetic in $(Fe_{4-x}Sn_x)N$ (see calculated values in figure 3c). Thus the magnetic structure of the $(Fe_{4-x}Sn_x)N$ alloys would have a strong tendency towards antiferromagnetic order with increasing Sn concentration. In figure 3d we show the calculated distance dependence of the of the 1a-1a exchange interactions in $(Sn_{0.9}Fe_{3.1})_4N$ alloys, which is the main subject of the present study. It can be seen that this interaction is quite short ranged and basically has only two significant members – 1NN and 2NN – both are antiferromagnetic which have similar magnitude. Already the 3NN interaction is one order of magnitude smaller, whereas more distant interactions have vanishing magnitude in the paramagnetic state.

Before we proceed with the analysis of the derived picture for the magnetic interactions in $(Sn_{0.9}Fe_{3.1})_4N$ some comments on the weak itinerant nature of magnetism of Fe in the 3c position in stoichiometric $Fe_4N$ and nearly stoichiometric $(Fe_{4-x}Sn_x)_4N$ are necessary. As it has been shown earlier the decrease of the Fe spin moments in the paramagnetic state compared to the ordered FM state due to their itinerant character is the leading mechanism of the anomalous negative thermal expansion in Fe-based Invar-like alloys via the mechanism of the spontaneous volume magnetostriction.[51,52] Indeed experimentally, the $Fe_4N$ is an Invar-like system[53] and thus the considerable reduction of the itinerant Fe(3c) moments in the paramagnetic state is completely in-line with this observation. A second comment is related to the situation of $(Fe_{4-x}Sn_x)_4N$ alloys around the critical concentration (x~0.2-0.3). It might appear that they would exhibit a very rich physics associated with the proximity to the onset of intrinsic magnetism on the Fe(3c) sublattice – a topic, which recently triggered considerable interest due to experimental observations of the "clean" effects of quantum criticality in disordered metallic alloys of $NiCoCr_x$, which can be driven towards an itinerant ferromagnetic critical point by chemical substitutions[54] - an effect predicted earlier for $CoGaNi_x$ and $CoGaNi_x$ alloys with similar structure.[55] It thus might be interesting to test experimentally the magnetic and critical behavior of the $(Fe_{4-x}Sn_x)_4N$ alloys around the critical compositions in the future.

Turning back to the $(Sn_{0.9}Fe_{3.1})_4N$ alloy one can summarize our discussion of the results presented in figure 3 as following. At this Sn concentration the Fe sites in the positions 3c do not possess an intrinsic magnetic moment – although some spin-polarization might be induced by the spontaneous magnetization of the Fe local moments in 1a positions. However, even this scenario of the induced spin-polarization in the 3c positions should be excluded since the moments on the 1a sites do not order ferromagnetically (thus it does not produce a molecular field on the 3c sites) since the dominant interactions within the Fe(1a) sub-lattice are antiferromagnetic. Thus the magnetism of $(Sn_{0.9}Fe_{3.1})_4N$ is due to well localized Fe moments randomly distributed over the simple cubic 1a sub-lattice, which populate 10% of sc lattice sites and interact antiferromagnetically such that the 1NN interactions are similar in magnitude to the 2NN ones (see figure 3d). This give us strong arguments pointing towards the notion that the experimentally observed[32] SG behavior in $(Sn_{0.9}Fe_{3.1})_4N$ is a special case of the SG phenomena in the random-site disordered systems on the non-geometrically frustrated sc lattice caused by the competition of the first two NN AFM couplings.

In order to further support our conclusions we perform a Monte-Carlo (MC) simulation with the Hamiltonian (1) and the calculated exchange constants as presented in figure 3d. The MC simulations have been done on 28x28x28 simple cubic lattice with 10% of sites randomly populated by magnetic atoms using periodic boundary conditions and a simple Metropolis algorithm. The classical magnetically ordered ground state on the ordered sc lattice (where all sites populated by magnetic atoms) with exchanges presented in figure 3d would be a collinear antiferromagnetic state[34] as shown in figure 4. This ordered state consists of ferromagnetic {001} chains with checkerboard like AFM ordering between in {001} planes of the sc lattice. In order to monitor the onset of magnetic order in disordered system we simulate the inter-site spin-spin correlation functions (CF) defined as for the $n^{th}$ nearest as

$$c(n) = \frac{1}{N}\sum_{i}\frac{1}{N_n}\sum_{\vec{R}_n}\langle \vec{e}_{\vec{R}_i}\vec{e}_{\vec{R}_i+\vec{R}_n}\rangle \,, \qquad (2)$$

where the first sum runs over $N$ translation vectors of the sc lattice, $\vec{R}_i$, the second sum is taken over the $N_n$ translation vectors, $\vec{R}_n$, spanning the $n^{th}$ shell, and $\langle\,\rangle$ stands for the statistical average. The spin-spin correlation function has been also normalized to take into the account the partial lattice site occupations by the magnetic atoms to give unity in the case of ferromagnetic order. For instance, with definition (2), the spin-spin correlation functions for fully ordered AFM structure from figure 4 will be c(1) = -0.33, c(2)=-1 and c(3)=+1. In figure 5 we plot the results of our MC simulations of CF for the first-four NN shells. As can be seen down to very low temperature (1K) no traces of the regular AFM order occurs. Due to rather significant values of the 1a-1a inter-site exchanges there is a strong short-range order (SRO) effect in the first two NN shells in a very broad temperature interval (up to hundreds K), but there is no trace of regular long range order (LRO) down to 1K. As temperature lowers c(1) and c(2) monotonously grow in magnitude without turning into the values they should have in an AFM state with LRO, merely increasing SRO defined by the respective signs of the exchanges between neighboring atoms in the respective shells. Quite notorious is the behavior of c(3) and c(4) – they also do not approach the AFM LRO values, but manifest a quite distinctive turning points at about 20K and 8K respectively, indicating an onset of strong cooperative behavior in this temperature interval, exactly where the onset of the spin glass behavior is observed experimentally[32] in $(Sn_{0.9}Fe_{3.1})_4N$.

In order to further elucidate the emergence of the SG state at the lowest temperature we calculate also the Edwards-Anderson (EA) parameter. The EA parameter is introduced here for the random sites model in an analogy with the random bonds model (Ref.[56]) as:

$$q_{EA} = \langle \vec{e}_i(t_0)\vec{e}_i(t_0 + t)\rangle \qquad (3),$$

where, similarly to equ. (2), the averaging stands for MC statistical average over all sites at a given temperature but in addition there is a time (*t*), average (over MC steps). This parameter has played an important role in SG theories based on the random bonds model and here, for a random-site model, it reflects the temporal evolution of the "magnetization" related to a single site. As one can see from figure 5 the EA parameters shows a clear upturn at a temperature around 15 K suggesting the onset of the temporal correlations of the moment directions on the single sites at this temperature. These temporal correlations are one of the important hallmarks of the SG behavior.

Our simulations with ab-initio calculated exchange constants thus predict that interactions, frustration, and site disorder prevent the formation of a state with magnetic LRO down to very low temperatures. This fact also points towards SG formation in $(Sn_{0.9}Fe_{3.1})_4N$ without geometrical frustration effects, long range interactions, and effects of cauterization. One needs to note, however, that the presented MC results, cannot be regarded as a proof of the SG state formation and that a more refined analysis beyond a simple Metropolis algorithm will be necessary. The idea of the present work is to direct the attention of theoreticians working on model simulations and spin glass theory to the recently discovered notorious SG system, $(Sn_{0.9}Fe_{3.1})_4N$ which, on basis of our ab-initio analysis and simple Monte-Carlo simulations, might be considered as one of the most "simple" realistic spin-glass materials.

**Figure captions:**

**Figure 1: Crystal structure of Fe4N. The Sn atoms populate Fe(1a) sites in the $(Fe_{4-x}Sn_x)_4N$ alloy. (color online)**

**Figure 2: Calculated inter-site exchange interactions in $Fe_4N$ compounds in the ferromagnetic (upper panel) and paramagnetic states with disordered local moments (lower panel). Closed symbols – interactions within the Fe(1a) sub-lattice, open – within the Fe(3c), and mixed = inter-sublattice - Fe(1a)-Fe(2c) interactions. The inter-atomic distance, d/a, is given in units of the cubic lattice constant a.**

**Figure 3: a) Calculated Fe magnetic moments in $(Fe_{4-x}Sn_x)_4N$ alloys in the ferromagnetic (red) and the disordered local moment states (dark) for two sub-lattices 1a – closed symbols, and 2c – open symbols; b) First nearest neighbor interaction between Fe(1a) and Fe(2c)calculated for $(Fe_{4-x}Sn_x)_4N$ alloys for ferromagnetic (circles) and disordered local moment state (squares): c) first (squares) and second (circles) nearest neighbor interactions between localized Fe in the position 1a calculated for $(Fe_{4-x}Sn_x)_4N$ alloys; c) The exchange interactions between Fe(1a) positions calculated for $(Sn_{0.9}Fe_{3.1})_4N$ alloy. (color online).**

**Figure 4, Collinear antiferromagnetic state on a simple cubic lattice -the ground state of the classical Heisenberg Hamiltonian on fully ordered lattice with interactions taken from figure 3d. (color online).**

**Figure 5.** Temperature dependence of the spin-spin correlation functions (equ. 2) for $(Sn_{0.9}Fe_{3.1})_4N$ alloy for the first four nearest neighbor shells and Edwards-Anderson order parameter, $q_{EA}$ (equ. 3). Results of the Monte-Carlo simulations with calculated exchange parameters from figure 3d (see text). (color online).

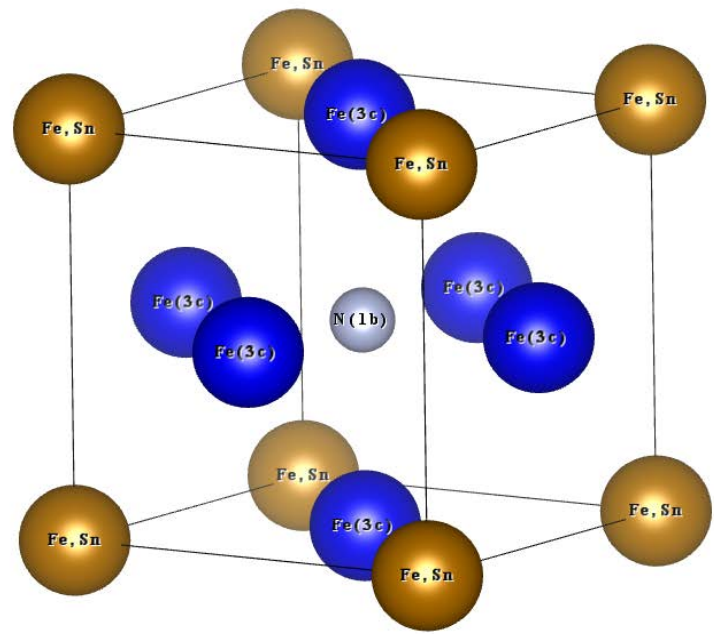

Figure 1.

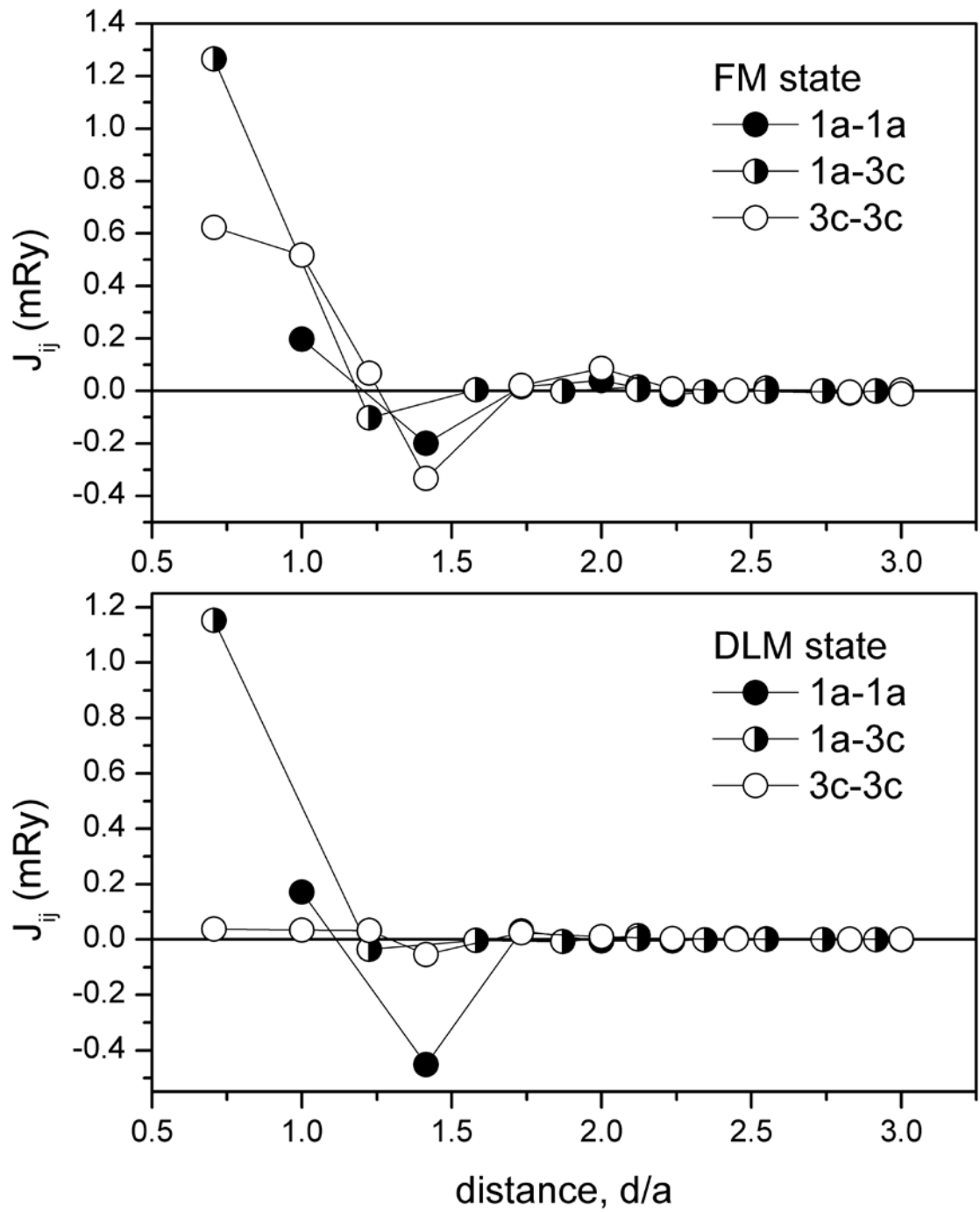

**Figure 2.**

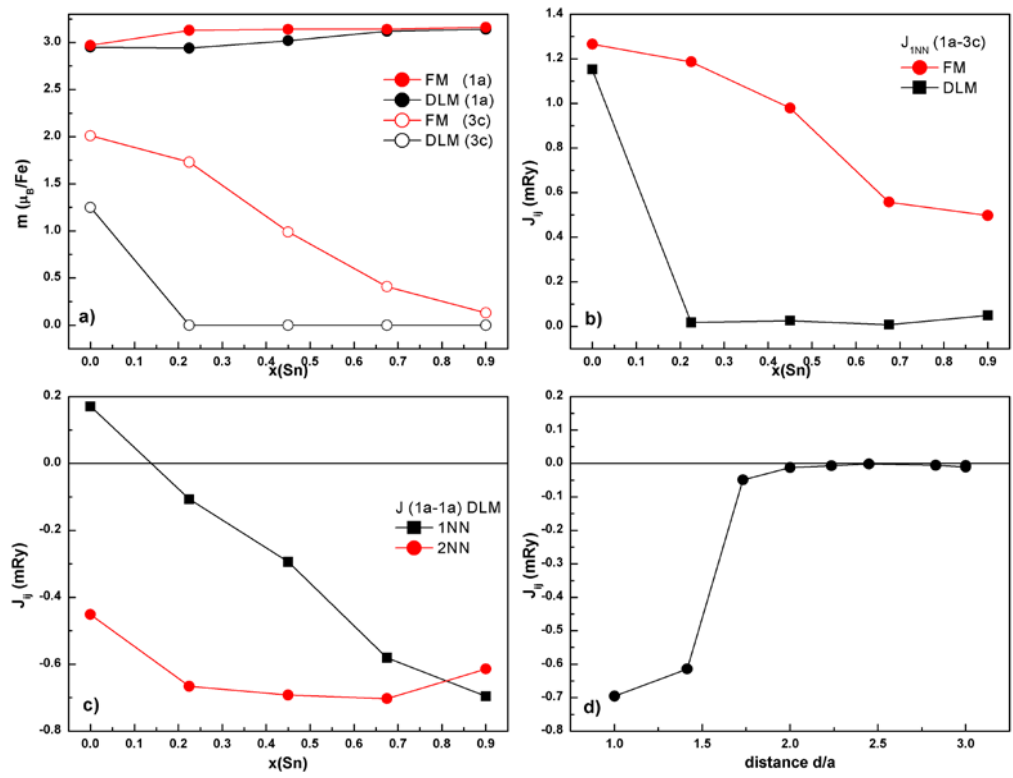

**Figure 3.**

**Figure 4.**

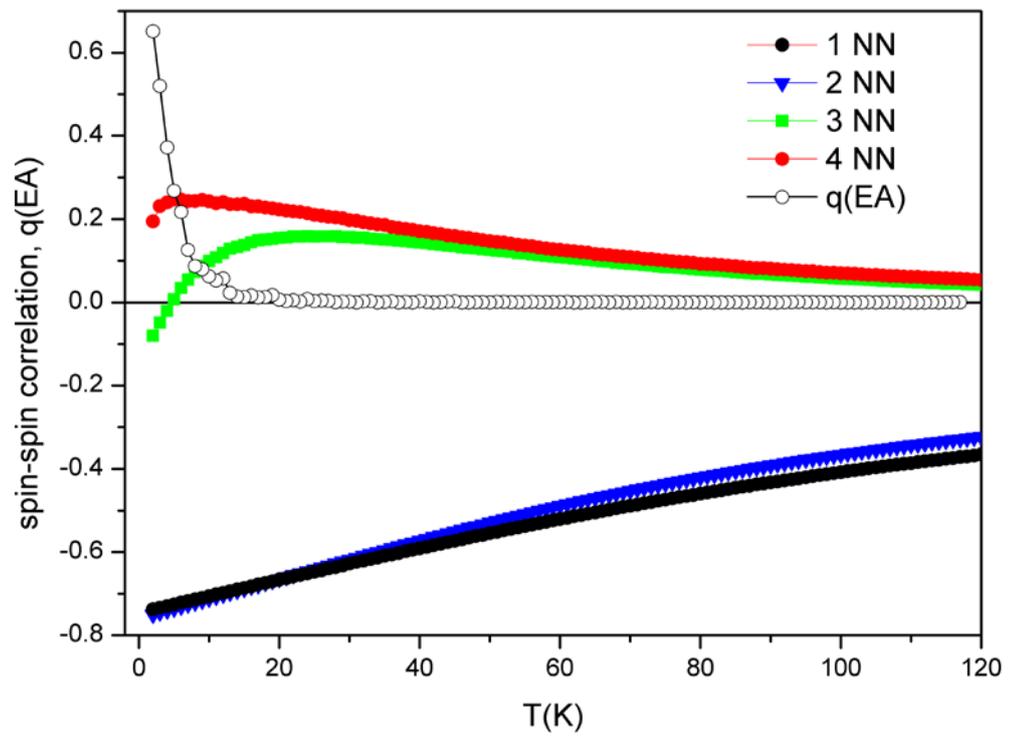

**Figure 5.**

# References.